\documentstyle[11pt,aaspp4]{article}
\newcommand{\vv}[1]{\mbox{\boldmath $#1$}}
\newcommand{\be}{\begin{equation}}
\newcommand{\ee}{\end{equation}}
\newcommand{\gapr}{\raisebox{-.6ex}{\mbox{
$\stackrel{>}{\mbox{\scriptsize$\sim$}}\:$}}}
\newcommand{\lapr}{\raisebox{-.6ex}{\mbox{
$\stackrel{<}{\mbox{\scriptsize$\sim$}}\:$}}}

\newcommand{\Tef}{T_{\rm eff}}

\newcommand{\Io}{I_{o}}
\newcommand{\Ie}{I_{e}}
\newcommand{\Fo}{F_{o}}
\newcommand{\Fe}{F_{e}}

\begin{document}
%%%%%%%%%%%%%%%%%%%%%%%%%%%%%%%%%%%%%%%%%%%%%%%%%%%%%%%%%%%%%%%%%%%%%
\title{Polarization of Thermal X-rays from Isolated Neutron Stars}
\author{G.~G. Pavlov}
\affil{The Pennsylvania State University, 525 Davey Lab,
University Park, PA 16802, USA; pavlov@astro.psu.edu}
\and
\author{V.~E. Zavlin}
\affil{Max-Planck-Institut f\"ur Extraterrestrische Physik, D-85740
Garching, Germany; zavlin@mpe.mpg.de}
%%%%%%%%%%%%%%%%%%%%%%%%%%%%%%%%%%%%%%%%%%%%%%%%%%%%%%%%%%%%%%%%%%%%%%
\begin{abstract}
Since the opacity of a magnetized plasma depends on polarization of radiation,
the radiation emergent from atmospheres of neutron stars with strong
magnetic fields is expected to be strongly polarized. The degree of linear 
polarization, typically $\sim 10-30\%$, depends on  photon energy,
effective temperature and magnetic field.  The spectrum of polarization 
is more sensitive to the magnetic field than the spectrum of intensity.
Both the degree of polarization and the position angle vary with the 
neutron star rotation period so that the shape of polarization pulse profiles 
depends on the orientation of the rotational and magnetic axes. Moreover, 
as the polarization is substantially modified by the general relativistic 
effects, observations of polarization of X-ray radiation from isolated 
neutron stars provide a new method for evaluating the mass-to-radius ratio 
of these objects, which is particularly important for elucidating the 
properties of the superdense matter in the neutron star interiors. 
\end{abstract}
%%%%%%%%%%%%%%%%%%%%%%%%%%%%%%%%%%%%%%%%%%%%%%%%%%%%%%%%%%%%%%%%%%%%
%%%%%%%%%%%%%%%%%%%%%%%%%%%%%%%%%%%%%%%%%%%%%%%%%%%%%%%%%%%%%%%%%%%%%%
\section{Introduction}
Optical and radio polarimetry has proven to be a powerful tool 
to elucidate properties of various astrophysical objects. For instance,
virtually all our knowledge about the orientations of the magnetic and
rotation axes of radio pulsars comes from analyzing the swing
of polarization position angle within the pulse (see, e.g., Manchester
\& Taylor 1977; Lyne \& Manchester 1988). On the other hand, X-ray polarimetry
has remained an underdeveloped field of astrophysics. Although various X-ray
polarimeters have been designed (e.g., Kaaret et al.~1990; 
Weisskopf et al.~1994; Elsner et al.~1997; Marshall et al.~1998), 
and importance of X-ray polarimetry convincingly demonstrated 
(M\'esz\'aros et al.~1988), most recent measurements of X-ray polarization 
has been made as long ago as in 1977, with the {\sl OSO 8} mission 
(Weisskopf et al.~1978). Nevertheless, it is expected that X-ray polarimeters 
will be launched in near future (see, e.g., Tomsick et al.~1997). To develop 
efficient observational programs for forthcoming X-ray missions whose 
objectives will include X-ray polarimetry, the problem of polarization of 
various X-ray sources should be carefully analyzed, with the main emphasis 
on new astrophysical information to be inferred from such observations.

In the present paper we consider polarization of thermal X-ray
radiation from isolated neutron stars (NSs) with strong magnetic fields.
Recent observations with the {\sl ROSAT} and {\sl ASCA} missions
have shown that several such objects are sufficiently bright for 
polarimetric observations --- e.g., the radio pulsars PSR 0833--45 and
PSR 0656+14 (see \"Ogelman 1995, and Becker \& Tr\"umper 1997, for reviews), 
and the radio-quiet NSs RX J0822--4300 (Zavlin, Tr\"umper, \& Pavlov 1999)
and RX~J1856.5--3754 (Walter, Wolk, \& Neuh\"auser 1996). Their soft X-ray 
radiation was interpreted as emitted from NS surface layers (atmospheres)
with effective temperatures $\Tef$ in the range of $(0.3-3)\times 10^6$~K. 
Since photons of different energies escape from different depths of the 
NS atmosphere with temperature growing inward, the spectrum of the thermal 
radiation may substantially deviate from the blackbody spectrum (Pavlov \& 
Shibanov 1978). Moreover, if there is a strong magnetic field in the 
NS atmosphere, such that the electron cyclotron energy $E_{Be}=\hbar eB/m_ec =
11.6\, (B/10^{12}~{\rm G})$~keV is comparable to or exceeds the photon
energy $E$, then the radiation propagates as two normal modes (NMs)
with different (approximately orthogonal) polarizations and 
opacities (Gnedin \& Pavlov 1974). For typical magnetic fields, 
$B\sim 10^{11}$--$10^{13}$~G, the NMs at soft X-ray energies, $E \ll E_{Be}$, 
are linearly polarized in a broad range of wavevector directions, 
and the opacity $\kappa_e$ of the so-called extraordinary mode
(polarized perpendicular to $\vv{B}$) is much smaller than that 
of the ordinary mode, $\kappa_e\sim (E/E_{Be})^2 \kappa_o$.
As a result, the extraordinary mode escapes from deeper and hotter layers,
so that the emergent radiation acquires strong linear polarization
perpendicular to the local magnetic field (Pavlov \& Shibanov 1978). 
Polarization of the observed radiation depends on the distribution of 
magnetic field and temperature over the visible NS surface.
If these distributions are axisymmetric, the polarization is a function of
the angle $\Theta$ between the symmetry (magnetic) axis 
and the line of sight. If the direction of the magnetic axis varies due to
NS rotation, the polarization patterns show pulsations with the period of 
rotation, so that measuring the polarization pulse profile allows one to 
constrain the orientations of the axes. Due to the gravitational bending of
the photon trajectories, the visible fraction of the NS surface
grows with increasing the NS mass-to-radius ratio, $M/R$, which reduces the net
polarization because the observer sees additional regions with differently
directed magnetic fields. On the other hand, the gravitational field affects
the magnetic field geometry making the field more tangential (Ginzburg \& 
Ozernoy 1965), which increases the observed polarization.
These GR effects enable one, in principle, to constrain $M/R$ by measuring
the X-ray polarization. We demonstrate that the expected X-ray
polarization of the thermal NS radiation is high enough to be measured
with soft-X-ray polarimeters in a modest exposure time, and these
measurements can provide important new information on both the geometry
of the magnetic field and the mass-to-radius ratio.

%%%%%%%%%%%%%%%%%%%%%%%%%%%%%%%%%%%%%%%%%%%%%%%%%%%%%%%%%%%%%%%%%%%%%%%
\section{Description of calculations}
%%%%%%%%%%%%%%%%%%%%%%%%%%%%%%%%%%%%%%%%%%%%%%%%%%%%%%%%%%%%%%%%%%%%%%%
The intensity $I$ and the Stokes parameters $Q$ and $U$
at a given point of the NS surface can be expressed 
as (Gnedin \& Pavlov 1974)
\begin{eqnarray}
I & = & \Io + \Ie~, 
\label{Iloc} \\
Q & = &  (\Io-\Ie)\,p_L\, \cos 2\chi_o~, 
\label{Qloc} \\
U & = &  (\Io-\Ie)\,p_L\,  \sin 2\chi_o~,
\label{Uloc}
\end{eqnarray}
where $\Io$ and $\Ie$ are the intensities of the ordinary and extraordinary
modes, $p_L=(1-{\cal P}^2)/(1+{\cal P}^2)$ is the degree of linear
polarization of the NMs (${\cal P}$ is the ellipticity, i.e.,
the ratio of the minor axis to the major axis of the polarization ellipse), 
and $\chi_o$ is the angle between the major axis of the polarization ellipse 
of the ordinary mode and the $x$ axis of a reference frame in which the 
Stokes parameters are defined.

We calculate the local NM intensities with the aid of NS atmosphere
models (e.g., Pavlov et al.~1994, 1995). In the present work we assume
that the surface temperature is high enough for the atmospheric matter
to be completely ionized. If the NS surface is covered with a hydrogen 
atmosphere, this assumption is justified at $\Tef \gapr 10^6$~K,
for typical magnetic fields of NS (Shibanov et al.~1993). The local 
intensities $\Io$ and $\Ie$ depend on the photon energy, magnetic field, 
and direction of emission.

In the dipole approximation, valid at photon and particle energies
much lower than $m_ec^2$, the degree of linear polarization of NMs can be 
expressed as
\be
p_L = 
\frac{|q|\,\sin^2\theta'}{\sqrt{4\cos^2\theta'+q^2\,\sin^4\theta'}}
\label{pL}
\ee
where $\theta'$ is the angle between the magnetic field $\vv{B}$ and the 
unit wavevector $\hat{\vv{k}}'$ at the NS surface. The (angle-independent) 
parameter $q$ is determined by the components of the Hermitian part of the 
polarizability tensor in the coordinate frame with the polar axis along the 
magnetic field (Gnedin \& Pavlov 1974). The parameter $q$ depends on photon 
energy and magnetic field (e.g., Pavlov, Shibanov, \& Yakovlev 1980; Bulik \& 
Pavlov 1996).  For instance, if the hydrogen plasma is completely ionized, and
the electron-positron vacuum polarization by the magnetic field can be 
neglected, this parameter equals
\be
q=\frac{E'^2(E_{Be}^2+E_{Bi}^2-E_{Be}E_{Bi})-E_{Be}^2E_{Bi}^2}
{E'^3(E_{Be}-E_{Bi})}~,
\label{q-param}
\ee
where $E'$ is the photon energy as measured at the NS surface,
$E_{Bi}=(m_e/m_p)E_{Be}=6.32\, (B/10^{12}~{\rm G})$~eV
is the ion (proton) cyclotron energy.  If the photon energy is much greater 
than the ion cyclotron energy, the $q$ parameter is particularly simple:
$q=E_{Be}/E'$.  This means that the NMs are linearly polarized, $p_L\simeq 1$,
in a broad range of directions, $\sin^2\theta' \gg 2E'/E_{Be}$, 
at photon energies much lower than the electron cyclotron energy.
It should be mentioned that equations (\ref{Iloc})--(\ref{pL})
imply that the NM polarizations are orthogonal to each other
(in particular, $\chi_e=\chi_o\pm \pi/2$).  This condition is fulfilled in a 
broad domain of photon energies and directions, except for a few special 
values of $\theta',E'$ (e.g., Pavlov \& Shibanov 1979; Bulik \& Pavlov 1996).
Within the same approximations, the angle $\chi_o$ coincides with the azimuthal angle of the magnetic field in a reference frame whose polar axis is parallel
to $\hat{\vv{k}}'$.

To find the observed flux $F_I$ and the observed Stokes parameters $F_Q$ and 
$F_U$, one should sum contributions from all the elements of the visible 
NS surface. We assume the magnetic field and the temperature distribution 
are axially symmetric and define $F_Q$ and $F_U$ in the reference frame such 
that the axis of symmetry $\vv{m}$ lies in the $xz$ plane, the $z$ axis is 
directed along the line of sight. In such a frame $F_U=0$, and $F_I$, $F_Q$ 
are functions of the angle $\Theta$ between $\vv{m}$
and $\hat{\vv{z}}$. The ratio $P_L=-F_Q/F_I$ gives 
the observed degree of linear polarization, $|P_L|$,
and the observed position angle: the polarization direction is perpendicular 
or parallel to the projection of $\hat{\vv{m}}$ onto the sky plane for
$F_Q>0$ or $F_Q<0$, respectively.

Since the NS radius $R$ is comparable with the gravitational
(Schwarzschild) radius $R_g=2GM/c^2$, the photon energy and 
the wavevector and polarization directions
change in the course of propagation
in the strong gravitational field. 
We will assume that the NS gravitational field is described by the exterior
Schwarzschild solution. Since strong magnetic fields
($B\gg 10^{10}$~G) are needed to
obtain measurable polarization in the X-ray range,
and all observed NSs with strong magnetic fields are not very fast rotators
($P>10$~ms), the effects of rotation on the metric and on the observed
radiation are very small. For the Schwarzshild metric,
the observed energy is redshifted as
$E=g_rE'$, where $g_r=(1-R_g/R)^{1/2}$ is the redshift factor. 
The observed wavevector is inclined to the emitted wavevector by the
angle $K-\vartheta$, $\hat{\vv{k}}'\vv{\cdot}\hat{\vv{z}}=\cos (K-\vartheta)$,
where $K$ is the colatitude of the emitting point in the reference
frame with the origin at the NS center and the $z$ axis directed towards
the observer, $\vartheta$ is the angle between the normal to the
surface and the wavevector direction $\hat{\vv{k}}'$
at the emitting point.  The angle $\vartheta$ would coincide
with $K$ in flat space-time. In the Schwarzschild geometry,
$K$ always exceeds $\vartheta$, i.e., some part of the NS back hemisphere
is visible. For instance,
\be
K = a \int_0^{R_g/R} \frac{{\rm d}x}{\sqrt{1-a^2(1-x)x^2}}~,
\label{Kcolat}
\ee
for $K\leq\pi$ ($g_r\geq 0.65$), where
$a=R\sin\vartheta/(R_g g_r)$ is the impact parameter in units of $R_g$
(see, e.g., Zavlin, Shibanov, \& Pavlov 1995).
In particular,
\be
K-\vartheta \simeq  u \tan\frac{\vartheta}{2}
+\frac{1}{16}u^2
\left[\frac{15(\vartheta -\sin\vartheta)}{\sin^2\vartheta} +
7\tan\frac{\vartheta}{2}\right]~
\ee
for $u\equiv R_g/R\ll 1$.

The bending of the photon trajectories is associated with changing
the direction of linear polarization. The polarization direction rotates 
in such a way that it keeps fixed orientation
with respect to the normal to the trajectory plane
(e.g., Pineault 1977), remaining perpendicular to the wavevector.
Without this rotation, the angles $\chi_o$ and $\chi_e$ would 
be conserved: $\chi_o^{\rm obs}=\phi$ at the observation point,
where $\phi=\tan^{-1}(B_y/B_x)$ is the azimuthal angle of the magnetic field
at the emitting point in the $x,y,z$ frame. To find $\chi_o^{\rm obs}$ 
with allowance for the GR effects, it is convenient to introduce the frame 
$x',y',z'$ such that the $z'$ axis is parallel to $\hat{\vv{k}}'$,
and the photon trajectory is in the $x'z'$ plane (see Fig.~1). The unit vectors along 
the axes of the two frames are connected with each other as follows
\begin{eqnarray}
\hat{\vv{x}}' & = & \cos\varphi\, \cos(K-\vartheta)\,\, \hat{\vv{x}}
+\sin\varphi\, \cos(K-\vartheta)\,\,
\hat{\vv{y}} -\sin(K-\vartheta)\,\, \hat{\vv{z}}~, \label{x-prime} \\
\hat{\vv{y}}' & = & -\sin\varphi\,\, \hat{\vv{x}} +\cos\varphi\,\, \hat{\vv{y}}~, \label{y-prime} \\
\hat{\vv{z}}' & = & \cos\varphi\, \sin(K-\vartheta)\,\, \hat{\vv{x}} +
\sin\varphi\, \sin(K-\vartheta)\,\,
\hat{\vv{y}} + \cos(K-\vartheta)\,\, \hat{\vv{z}}~,
\label{z-prime}
\end{eqnarray}
where $\varphi$ is the azimuthal angle of the emitting point in the
$x,y,z$ frame. Using the conditions that the angle between $\hat{\vv{y}}'$
and the polarization direction is conserved, and the polarization direction
is perpendicular to $\hat{\vv{z}}$ at the observation point, we obtain
$\chi_o^{\rm obs} = \varphi +\phi'$, where $\phi'$ is the azimuthal angle 
of the magnetic field in the $x',y',z'$ frame: 
$\phi'=\tan^{-1}(B_{y'}/B_{x'})$. The angle $\phi'$ depends on
$K-\vartheta$ and $\varphi$, and it tends to $\phi-\varphi$ when $R_g/R\to 0$.

With allowance for the above-described gravitational effects,
the observed flux $F_I$ and the Stokes parameter $F_Q$ are given
by the following integrals over the visible NS surface 
(see Zavlin et al.~1995):
\begin{eqnarray}
F_I(E,\Theta) & = &
\frac{R^2}{d^2} g_r \int_0^1 \mu\, 
{\rm d}\mu \int_0^{2\pi} {\rm d}\varphi\, 
(\Io+\Ie)\, = \, \Fo(E,\Theta) + \Fe(E,\Theta)~,
\label{netI} \\
F_Q(E,\Theta) & = & \frac{R^2}{d^2} g_r \int_0^1 \mu\, 
{\rm d}\mu \int_0^{2\pi} {\rm d}\varphi\,
(\Io - \Ie)\, p_L\, \cos2(\varphi+\phi')~,
\label{netQ}
\end{eqnarray}
where $d$ is the distance, $\mu=\cos\vartheta$, and the integrands
are taken at the photon energy $E'=E/g_r$.

To calculate the integrands, we should know the magnetic field
at the NS surface as a function of $\mu$ and $\varphi$.
We consider a dipole magnetic field in the Schwarzschild metric.
According to Ginzburg \& Ozernoy (1965), the field equals
\be
\vv{B} = B_p \frac{(2+f)(\hat{\vv{r}}\cdot\hat{\vv{m}})\hat{\vv{r}}
-f\hat{\vv{m}}}{2}~,
\label{Bloc}
\ee 
where $B_p$ is the field strength at the magnetic pole, $\hat{\vv{r}}$ is
the unit radius vector of a surface point, and 
$\hat{\vv{m}}$ is the unit vector of the magnetic moment. The parameter
\be
f = 2 \frac{u^2-2u-2(1-u)\ln(1-u)}{[u^2+2u+2\ln(1-u)]\sqrt{1-u}}~
\label{f-effect}
\ee
accounts for the GR effect. For $u\ll 1$ ($R\gg R_g$), we have
%\be
%f(u) \simeq 1+\frac{1}{4}u+\frac{11}{80}u^2~.
%\ee
%
$f(u) \simeq 1+ u/4+11u^2/80$.
The radial and tangential components
of the magnetic field are $B_r=B_p\cos\gamma$ and $B_t=(B_p/2)f\sin\gamma$,
where $\cos\gamma=\hat{\vv{r}}\vv{\cdot}\hat{\vv{m}}=
\sin\Theta\sin K\cos\varphi + \cos\Theta\cos K$. Since $f>1$,
the GR effect makes the magnetic field more tangential.
The projections of $\vv{B}$ onto the $x,y,z$ and $x',y',z'$
axes can be easily found with the aid of 
equations 
\begin{eqnarray}
\hat{\vv{r}} & = & \sin K\, \cos\varphi\,\, \hat{\vv{x}}+
\sin K\, \sin\varphi\,\, \hat{\vv{y}}+
\cos K\, \hat{\vv{z}}~,
\label{radvec} \\
\hat{\vv{m}} & = & \sin\Theta\,\, \hat{\vv{x}}+ \cos\Theta\,\, \hat{\vv{z}}~,
\label{magnmoment}
\end{eqnarray}
and equations (\ref{x-prime})--(\ref{z-prime}).
The strength of the magnetic field is
\be
B=(B_p/2)\left[(4-f^2)\,\cos^2\gamma +f^2\right]^{1/2}=
B_p f\left[4-(4-f^2)\,\cos^2\theta_B\right]^{-1/2}~,
\label{magnstrength}
\ee
where $\theta_B$ is the angle between $\vv{B}$ and the normal to the surface
$\hat{\vv{r}}$.

The integration over the NS surface (eqs.~[\ref{netI}], [\ref{netQ}])
proceeds as follows.  For each point of the $\mu,\varphi$ grid, 
we calculate the colatitude $K(\vartheta)$ from equation~(\ref{Kcolat})
and the components of the radius vector $\hat{\vv{r}}$ in the $x,y,z$ and
$x',y'z'$ frames (eqs.~[\ref{radvec}] and [\ref{x-prime}]--[\ref{z-prime}]).
This gives us the projections and strength of the local magnetic field $\vv{B}$
(eqs.~[\ref{Bloc}] and [\ref{magnstrength}]), the angles 
$\phi'=\tan^{-1}(B_{y'}/B_{x'})$, $\theta'=\cos^{-1}(B_{z'}/B)$,
and $\theta_B=\cos^{-1}(\vv{B}\vv{\cdot}\hat{\vv{r}})$,
and the degree of NM polarization $p_L$ (eq.~[\ref{pL}]).
To obtain the intensities of the extraordinary and ordinary modes of radiation
emitted to the observer from a given surface point, one needs to 
know the local depth dependences of the temperature and density in the NS
atmosphere, determined by the local values of $B$, 
$\theta_B$, and $T_{\rm eff}$.  These dependences
are obtained by interpolation within a set of 
the diffusion atmosphere models (Pavlov et al.~1995). 
Then, the $\Io$ and $\Ie$ intensities are computed as 
described by Pavlov et al.~(1994) and Shibanov \& Zavlin (1995).
Subsequent numerical integration over $\mu,\varphi$ gives us $F_I$, $F_Q$,
the degree of the observed linear polarization 
and the position angle.

%%%%%%%%%%%%%%%%%%%%%%%%%%%%%%%%%%%%%%%%%%%%%%%%%%%%%%%%%%%%%%%%%%%%%%%
\section{Results}
%%%%%%%%%%%%%%%%%%%%%%%%%%%%%%%%%%%%%%%%%%%%%%%%%%%%%%%%%%%%%%%%%%%%%%%
To demonstrate how the observed linear polarization depends on 
photon energy, magnetic field, and mass-to-radius ratio, 
we consider a NS covered with a hydrogen atmosphere with a uniform 
effective temperature and a dipole magnetic field. We present the results for
$T_{\rm eff}=1\times 10^6$~K, $B_p/(10^{12}~{\rm G})=0.3$, 1.0, 3.0 and 10.0. 
We choose a standard NS radius $R=10$~km and three NS masses,
$M/M_\odot=0.66$, 1.40 and 1.92, from an allowed domain in the $M$--$R$ 
diagram (filled circles in Fig.~2).  These masses correspond to
the redshift parameters $g_r=0.90$, 0.77 and 0.66,
and the surface gravitational accelerations 
$g/(10^{14}~{\rm cm}^2~{\rm s}^{-1}) =0.97$, 2.43, and 3.89.
It should be noted that the properties of the emitted radiation are almost
independent of the $g$ value, so that the gravitational effects on 
the observed radiation are determined mainly
by the redshift parameter $g_r$, i.e., by the mass-to-radius ratio. 

The left panel of Figure~3 demonstrates the observed photon spectral fluxes 
$F_I$ (eq.~[\ref{netI}]) from a NS with the magnetic field
$B_p=1\times 10^{12}$~G at the magnetic pole and the magnetic axis 
perpendicular to the line of sight, $\Theta=90^\circ$. The flux is normalized 
to a distance of 1~kpc. The spectra are presented for the three values of 
the redshift parameter $g_r$.  To demonstrate the effect of the interstellar
absorption, we plot the spectra for the effective hydrogen column
densities $n_H=0$ (unabsorbed flux), $1\times 10^{20}$
and $1\times 10^{21}$~cm$^{-2}$ (the latter two are shown for $g_r=0.77$ only).
The effect of redshift is clearly seen, as well as that of the interstellar 
absorption: spectral maxima shift from $0.15$~keV at $n_H=0$ to
$0.6$~keV at $n_H=1\times 10^{21}$~cm$^{-2}$.
The contributions from the extraordinary
and ordinary modes (fluxes $\Fe$ and $\Fo$)
to the unabsorbed spectrum $F_I=\Fe+\Fo$ 
are shown in the right panel of Figure 3 for $g_r=0.77$. 
At energies around the maxima of the flux spectra
the radiative opacity of the ordinary mode significantly exceeds
that of the extraordinary mode. Hence, the extraordinary mode is emitted
from deeper and hotter atmosphere layers, providing the main contribution
to the total flux (see Pavlov et al.~1995 for details). 
At higher energies ($E'\gapr 3^{-1/2} E_{Be}$)
the relation between the two opacities, and the two NM fluxes, is reversed 
(Kaminker, Pavlov, \& Shibanov~1982). This leads to changing the sign of 
$F_Q$ (i.e., to the jump of the polarization position angle by $\pi/2$).

Several examples of the dependences of $P_L=-F_Q/F_I$ on photon energy
are presented in Figure 4,
for $B_p=1\times 10^{12}$~G and different 
values of the angle $\Theta$ and the redshift parameter $g_r$.
In the soft X-ray range, where the thermal NS radiation is 
most easily observed, $P_L$ is positive,
i.e., the polarization direction is perpendicular to the projection
of the NS magnetic axis onto the image plane.
In this energy range the ordinary mode is 
emitted from superficial layers with lower temperature,
whereas the extraordinary mode is formed in deeper and hotter 
layers with a larger temperature gradient.
As a result, the ratio $\Fe/\Fo$ grows with $E$ at low energies
until this effect is compensated by the decrease of the difference
between the extraordinary and ordinary opacities 
(see the right panel of Fig.~3).
At higher energies the extraordinary and ordinary fluxes approach each
other, so that $\Fe/\Fo$ decreases with increasing $E$ reaching unity at
$E\approx 0.3 g_r E_{Be}$ (hereafter, $E_{Be}$ and $E_{Bi}$ are the cyclotron
energies for the magnetic field $B_p$).
Since $p_L\approx 1$ at $E_{Bi}\ll E'\ll E_{Be}$,
and $\varphi+\phi'$ does not depend on $E'$, it follows from equations
(\ref{netI}) and (\ref{netQ}) that $P_L\propto (\Fe-\Fo)/(\Fe+\Fo)$,
with a proportionality coefficient independent of $E$.
This explains the energy dependence of $P_L$ in Figure 4.
In particular, starting from energies $E\sim g_r E_{Bi}$, $P_L$ grows
with $E$ until it reaches a maximum (at $E \sim 1$~keV for $B_p=1\times
10^{12}$~G). At higher energies the polarization spectra steeply 
decrease with increasing $E$, 
reach zero at $E\sim 0.3 g_r E_{Be}$, where the contributions
from the two NMs cancel each other, and become negative (polarization   
direction becomes parallel to the $\vv{m}$ projection)
at higher energies, where the flux decreases exponentially at the
effective temperature chosen.

As expected, the degree of polarization grows with 
increasing $\Theta$ from $0^\circ$ to $90^\circ$.
It equals zero at $\Theta=0^\circ$ 
(the magnetic axis is parallel to the line of sight)
because the contributions from the azimuthal angles $\varphi$ and
$\varphi+\pi/2$ to $F_Q$ (eq.~[\ref{netQ}]) are polarized in orthogonal
directions, being of the same magnitude ($B_{y'}=0$, $\phi'=0$,
$\cos 2\varphi=-\cos 2(\varphi\pm \pi/2)$).
The polarization is maximal at $\Theta=90^\circ$, when 
the magnetic lines are seen by the observer almost parallel
to the magnetic axis on a substantial (central)
part of the visible stellar disk.

We see from Figure 4 that the degree of polarization decreases
with increasing $M/R$ (or decreasing $g_r$). The main reason is that
the observer sees a larger fraction of the whole NS surface
due to stronger  bending of photon trajectories.
As a result, the overall pattern of the magnetic lines on the visible
NS disk becomes more nonuniform, which leads to additional ``cancellation''
of the mutually orthogonal polarizations emitted from parts of the surface with
orthogonally oriented magnetic line projections. 
This effect is partly compensated by the other GR effect, more tangential
dipole magnetic field in the Schwarzschild metric (the parameter
$f$ in eq.~[\ref{Bloc}]
equals 1.08, 1.14, and 1.22 for $g_r=0.90$, 0.77, and 0.66,
respectively), but the effect of bending prevails.

Figure 5 demonstrates the effect of magnetic field strength on
the degree of polarization. In particular, $P_L$ in the soft X-ray range
grows with $B_p$ for typical NS magnetic fields. When the magnetic field is 
relatively small, $B_p\lapr 1\times 10^{12}$~G, the fast
growth of $P_L$ is due to the increasing difference
between the extraordinary and ordinary opacities (hence,
increasing differences between the escape depths
and between the NM intensities). Slower growth of $P_L$ at intermediate 
fields (around $B_p\sim 1\times 10^{12}$~G) is mainly due to the decrease
of the surface temperature with increasing $B$ (Pavlov et al.~1995),
which reduces the ordinary flux $\Fo$ and the ratio $\Fo/\Fe$.
With further increase of magnetic field, the surface temperature ceases 
to decrease, and $P_L(B_p)$ saturates in the 
soft X-ray range at $B_p\gapr 3\times 10^{12}$~G, 
until the field becomes so strong, $B_p\gapr 3\times
10^{13}$~G, that $g_rE_{Bi}$ becomes comparable with $E$, and the proton 
cyclotron spectral feature gets into the soft X-ray range.

The proton cyclotron feature in the polarization spectrum
is shown in Figure 5 for $B_p=1\times 10^{13}$~G.
The shape of the feature can be explained by the behavior
of the parameter $q$ and the NM intensities
near the proton cyclotron resonance. According to equation (\ref{q-param}),
$q$ grows with decreasing photon energy until $E'$ reaches $3^{1/2} E_{Bi}$;
then it sharply decreases, crosses zero in the very vicinity of 
proton cyclotron resonance, at $E'\simeq E_{Bi} (1+2m_e/m_p)$,
and tends to $-\infty$ ($q\simeq -E_{Be}E_{Bi}^2/E'^3$ at $E'\ll E_{Bi}$).
This means that $p_L$ reaches zero, the NM intensities $\Ie$ and $\Io$ 
equal each other, and the integrand of equation (\ref{netQ}) equals zero
at an energy in the very vicinity of the proton resonance
corresponding to the local magnetic field. 
If the direction of the local magnetic field is
such that $\cos 2(\varphi+\phi')>0$ (see eq.~[\ref{netQ}]),
which roughly corresponds to the projection of the local magnetic
field onto the sky plane within $\pm 45^\circ$ of the magnetic
axis projection, then the integrand in the expression for $-F_Q$ is 
positive at  energies around the resonance, so that the energy dependence 
of the integrand for the corresponding surface points looks like an 
``absorption line'' in a positive continuum, with its minimum (zero) 
value at the local resonance energy. Integration over the area with positive
$\cos 2(\varphi+\phi')$ yields a positive contribution to $-F_Q$,
with an absorption line somewhat broadened, and its minimum above zero, 
because of nonuniformity of the magnetic field.
On the contrary, the energy dependence of the integrand in the area
where $\cos 2(\varphi+\phi')<0$ looks like an ``emission line''
on a negative continuum, with its maximum equal zero at the
local resonance energy. The integral over this area gives a negative
contribution to $-F_Q$, its absolute values are minimal in the energy
range which includes the local resonance energies.
If the strengths of the local magnetic fields are different in
the areas of positive and negative $\cos 2(\varphi+\phi')$,
the integration over the whole visible disk
results in a complex feature in the $-F_Q(E)$ and $P_L(E)$ spectra.
In Figure 5 this feature is most pronounced at
$B_p=1\times 10^{13}$~G and $\Theta=90^\circ$. It consists
of two components: the absorption component with a sharp minimum
at $E\simeq 28$~eV corresponding to the field at the magnetic equator 
($B_e=B_pf/2$), and the emission component at energies below 
$E\simeq 48$~eV, corresponding to the field at the magnetic pole.
The feature is also clearly seen at the same $B_p$ and
$\Theta=45^\circ$, whereas only the emission component of the feature
is seen at $E>10$~eV in the curves of $P_L(E)$ for $B_p=3\times 10^{12}$~G.
If the magnetic field is superstrong, $B\sim 10^{14}-10^{15}$~G,
as suggested for anomalous X-ray pulsars and soft gamma repeaters,
the proton cyclotron feature gets into the soft or medium X-ray
range, and its detection would enable one to measure directly
the magnetic field strength.

Generally, the polarization spectra are much more 
sensitive to the strength of magnetic field than the intensity spectra
(Fig.~6). The main effect of $B$ on $F_I(E)$ spectra 
in the soft energy range is a shallow proton cyclotron absorption feature
(see the spectrum for $B_p=1\times 10^{13}$~G in
Fig.~6). At $E\sim 0.2-1.0$~keV the intensity spectra
at different magnetic fields typical for NSs are almost
indistinguishable, with the main contribution coming from the 
extraordinary mode whose spectrum is almost independent of $B$
at $E_{Bi}\ll E'\ll E_{Be}$.

If the angle $\alpha$ between the magnetic and rotation axes differs
from zero, the projection of $\vv{m}$ onto the sky changes its orientation
with the period of rotation. This means that the angle $\Theta$
and the polarization position angle $\delta$ with respect
to a fixed (nonrotating) direction also oscillate with the rotation period
$P$:
\be
\cos\Theta = \cos\zeta\,\cos\alpha + \sin\zeta\,\sin\alpha\,\cos 2\pi\Phi~,
\label{Theta}
\ee
\be
\tan\delta =  \frac{\cos\alpha\,\sin\zeta - \sin\alpha\,\cos\zeta\, \cos 2\pi\Phi}
{\sin\alpha\, \sin 2\pi\Phi}~,
\label{posangle}
\ee
where $\zeta$ is the angle between $\vv{\Omega}$ 
(NS rotation axis) and the line of sight,
$\Phi =t/P$ is the rotation phase, the angle $\delta$ is counted from 
the projection of $\vv{\Omega}$ onto the sky. Equation (\ref{posangle}) 
is applicable in the case of polarization perpendicular to the projection 
of $\vv{m}$ onto the sky plane ($P_L>0$); for $P_L<0$, the left-hand-side 
is replaced by $\cot\delta$. 

Figure 7 shows several characteristic dependences of $P_L(\Phi)$ and
$\delta(\Phi)$ for $E=0.3$~keV, $g_r=0.77$, and a few sets of $\zeta,\alpha$. 
In the particular case of an orthogonal rotator,
$\zeta=\alpha=90^\circ$, we have $\Theta=\Phi$, $\delta=0$, i.e.,
the degree of polarization oscillates between zero (at $\Phi=0$, 0.5)
and a maximum value, $P_L=25\%$ at $\Phi=0.25$, 0.75, showing two pulses per
period, while the position angle remains constant
(the polarization is oriented along the direction of the
rotation axis). For the case $\zeta=60^\circ$,
$\alpha=50^\circ$, the minimum polarization, $P_L=1\%$ at $\Phi=0$, 1,
corresponds to $\Theta=\zeta-\alpha=10^\circ$. The polarization pulse
has two maxima per period, $P_L=25\%$ at $\Phi=0.36$, 0.64 (corresponding
to $\Theta=90^\circ$) and a local minimum, $P_L=22\%$ at $\Phi=0.5$
(corresponding to $\Theta=\zeta+\alpha=110^\circ$). The position angle
oscillates around $\pi/2$ (or $-\pi/2$). For
$\zeta=\alpha=45^\circ$, $P_L$ has one broad maximum per period,
of the same height as for the orthogonal rotator, when $\Theta=90^\circ$
at $\Phi=0.5$. The position angle swings from 0 to
$\pi$ during the period,
crossing $\pi/2$ at the phases of maximum polarization.
Finally, for $\zeta=40^\circ$, $\alpha=10^\circ$, the phase dependence of
the degree of polarization is almost sinusoidal, with one maximum per period;
it oscillates between $7\%$ and $16\%$.
The position angle oscillates around $\pi/2$ (or $-\pi/2$) with a small
amplitude because of the small value of $\alpha$.
Thus, we see that a variety of the phase dependences of $P_L$
and $\delta$ can be obtained for various combinations of $\zeta,\alpha$,
which is potentially useful for evaluating these angles from polarimetric
observations.

%%%%%%%%%%%%%%%%%%%%%%%%%%%%%%%%%%%%%%%%%%%%%%%%%%%%%%%%%%%%%%%%%%%%%5
\section{Discussion}
%%%%%%%%%%%%%%%%%%%%%%%%%%%%%%%%%%%%%%%%%%%%%%%%%%%%%%%%%%%%%%%%%%%%%55
As we see from the examples above, the degree of linear polarization
of thermal NS radiation is
quite high, up to 20\%--50\% in pulse peaks, and about twice lower
for the phase-averaged polarization, for typical NS parameters.
Optimal energies for observing the polarization are in the soft
X-ray range, $\sim 0.1$--1~keV, for typical surface temperatures
of young and middle-aged NSs.
The polarization can be observed at these energies with the use of multilayer
coated mirrors which provide high reflectivity at large grazing
angles --- see Marshall et al.~(1998) for a concept for a satellite-borne
polarimeter (PLEXAS) which would be able to measure the linear polarization
from brightest thermally emitting pulsars to an accuracy of 1\%--3\%
during modest exposure times $\sim 30$--100~ks. 
The polarization can be measured in several narrow energy bands.  
Complementing the spectral flux with 
the polarization spectral data and comparing the both with the NS
atmosphere models would considerably narrow allowed ranges for NS
parameters such as the magnetic field and effective temperature.
Particularly strong constraints could be obtained if the X-ray polarimetric
measurements are supplemented by measuring optical-UV polarization
of the same sources. Several middle-age pulsars have been observed
successfully in the optical-UV range with the {\sl Hubble Space Telescope}
(e.g., Pavlov, Welty, \& C\'ordova 1997, and references therein).
Although the expected polarization in this range is lower than in soft
X-rays, it still can be as high as 5\%--15\% (see Figures 4 and 5),
so that UV-optical polarimetric observations of these sources seem
quite feasible.

Using detectors with high timing resolution (e.g., microchannel-based
photon counters) for polarimetric observations of pulsars would allow
one to measure phase dependences of the degree of polarization and
the position angle. These observations would be most useful to
determine the 
inclinations of the rotation and magnetic axis ($\zeta$ and $\alpha$).
Although radio band polarization data have been
widely used to constrain these angles, 
the results are often ambiguous because the same behavior of the position
angle can be fitted with different combinations of $\zeta$ and
$\alpha$. An additional difficulty in interpreting the radio
polarization data is caused by shortness of the duty cycle of radio
pulsars --- the radio flux is too low during a substantial fraction
of the period to measure the polarization. Since the X-ray flux
remains bright during the whole period, measuring the position
angle and the degree of polarization in X-rays
would enable one to infer the orientation of the pulsar axes
with much greater certainty.

Of particular interest is the result that the observed polarization
is sensitive to the NS mass-to-radius ratio, the most crucial parameter
to constrain still poorly known equation of state of 
the superdense matter in the
NS interiors. Since the radio emission of pulsars is generated well
above the NS surface, this ratio 
cannot be constrained from polarimetric observations
of pulsars in the radio band. Although some constraints can be obtained
from the pulse profiles of the X-ray flux (Pavlov \& Zavlin 1997),
the polarization pulse profiles are more sensitive to the 
gravitational effects.

Primary targets for studying the X-ray polarization of thermal
NS radiation are the X-ray brightest, thermally radiating
pulsars such as PSR 0833--45 (Vela) and PSR 0656+14.
A prototype of another class of promising targets, radio-quiet NSs
in supernova remnants, is RX~J0822--4300 in Puppis A. 
X-ray polarimetric 
measurements would be crucial to establish the strength and geometry
of the magnetic field of this putative isolated X-ray pulsar
(Pavlov, Zavlin, \& Tr\"umper 1999).
Also it would be very interesting to study X-ray polarization of
X-ray bright, radio-quiet isolated NSs which are not associated with
supernova remnants. Particularly interesting is the object
RX~J1856.5--3754 which has a thermal-like soft X-ray spectrum
but does not show pulsations (Walter et al.~1996). The lack 
of pulsations may be explained either by 
smallness of the magnetic
inclination $\alpha$, if the magnetic field is typical for NSs
($\sim 10^{11}$--$10^{13}$~G), or
by a very low surface magnetic field. Polarimetric
observations would enable one to distinguish between the two hypotheses
--- the polarization is expected to be high (and unpulsed)
in the former case (unless
$\zeta$ is also small), and it would be very low if the magnetic field
is lower than $\sim 10^{10}$~G. Distinguishing between the two options
is needed to choose either low-field or high-field NS atmosphere
models --- applications of these types of models to interpretation
of the multiwavelength observations of this source 
yield quite different NS parameters (Pavlov et al.~1996).

It follows from our results that thermal radiation from
millisecond pulsars, whose typical
magnetic fields are $10^8-10^9$~G ($E_{Be}\sim 1-10$~eV), 
is not polarized in the X-ray range. However, their polarization
is expected to be quite strong in the optical-UV range and can
be measured in sufficiently deep observations. The best candidate
for such observations is the nearest millisecond pulsar J0437--4715
whose magnetic field, $B\approx 8\times 10^8$~G, was estimated from
radio observations.
The thermal radiation from the NS surface, which is expected to be
heated up to $\sim 10^5$~K, should prevail over the radiation
from the very cool white dwarf companion at $\lambda \lapr 2000$~\AA.
This means that polarization from this pulsar could be observed
in the far-UV range with the {\sl Hubble Space Telescope}. It should be
mentioned that the rapid rotation of millisecond pulsars may affect
not only the intensity pulse shape (Braje, Romani, \& Rauch 1999),
but also the polarization, the effect neglected in the present paper.

In summary, our results demonstrate that including X-ray polarimeters
in future X-ray observatories, or launching dedicated X-ray polarimetry
missions, would be of great importance for studying NSs. The polarimetric
observations would be useful for studying not only the thermal component
of the NS X-ray radiation, but also the nonthermal component which
dominates in many X-ray emitting pulsars, particularly at higher energies.
Further theoretical investigation of polarization of
both thermal and nonthermal X-ray
emission from NSs are also warranted to provide firm interpretation
of future observational data. In particular, it would be useful to
consider the polarization with allowance for possible nonuniformity of
the temperature distribution over the NS surface and to study the effects
of chemical composition of the NS atmosphere on polarization.

%%%%%%%%%%%%%%%%%%%%%%%%%%%%%%%%%%%%%%%%%%%%%%%%%%%%%%%%%%%%%%%%%%%%%%%
\acknowledgements
We are grateful to Hermann Marshall for the useful discussions
of capabilities of modern soft X-ray polarimeters.
This work has been partially supported through NASA grant NAG5-7017.

%%%%%%%%%%%%%%%%%%%%%%%%%%%%%%%%%%%%%%%%%%%%%%%%%%%%%%%%%%%%%%%%%%%%%%%

\newpage
\figcaption{
Angles and vectors used in the paper. 
}

\figcaption{
Mass-radius diagram for NSs. The thick solid curves show the $M(R)$ dependences
for several equations of state of the superdence matter (Shapiro \& Teukolsky
1983): soft ($\pi$), intermediate (FP) and stiff (TI and MF). 
The thick straight line, $R=1.5 R_g$, corresponds to the
most conservative lower limit on NS radius at a given mass.
Thin dashed lines correspond to different values of the redshift  
parameter $g_r$ (the numbers near the lines). The filled circles
give the NS mass and radius used in our computations.
}

\figcaption{
{\sl Left}: 
Photon spectral fluxes $F_I(E)$ for $d=1$~kpc, $B_p=1\times
10^{12}$~G, $\Theta=90^\circ$,
and three values of the redshift parameter, $g_r=0.66$, 
0.77, and 0.90 (dash-dot, solid, and dashed curves, respectively).
The numbers near the curves denote the interstellar hydrogen column
density $n_H$, in units of $10^{20}~{\rm cm}^{-2}$.
{\sl Right}: Contributions from the extraordinary ($e$) 
and ordinary ($o$) NMs to the observed flux for $g_r=0.77$, $n_H=0$.
}

\figcaption{
Effect of the mass-to-radius ratio and the
inclination $\Theta$ of the magnetic axis on 
the polarization spectrum for $B_p=1\times 10^{12}$~G.
Thin, medium, and thick curves are for
$g_r=0.90$, 0.77, and 0.66, respectively.
}

\figcaption{
Effect of the strength of magnetic field on the polarization spectrum
for $g_r=0.77$.
The curves are plotted for $\Theta=90^\circ$ (solid) and $45^\circ$ (dashed),
and $B_p/(10^{12}~{\rm G})=0.3$, 1.0, 3.0, and 10.0 
(the numbers near the curves).
}

\figcaption{
Unabsorbed spectral fluxes $F_I(E)$ for $g_r=0.77$ and 
$B_p/10^{12}~{\rm G} = 0.3$ (short dashes), 1.0 (solid), 
3.0 (long dashes), and 10.0 (dash-dots).
}

\figcaption{
Dependences of the degree of polarization and the position angle on
NS rotation phase $\Phi$ for $g_r=0.77$, $E=0.3$~keV, and different
angles between the rotaiton axis and the line of sight, and between
the magnetic and rotation axes:
$\zeta=\alpha=90^\circ$ (solid), $\zeta=\alpha=45^\circ$ 
(long dashes), $\zeta=60^\circ$ and $\alpha=50^\circ$ (dot-dashes),
and $\zeta=40^\circ$ and $\alpha=10^\circ$ (short dashes).
}


\begin{references}
\reference{}
Becker, W., \& Tr\"umper, J. 1997, A\&A, 326, 628

\reference{}
Braje, T.~M., Romani, R.~W., \& Rauch, K.~P. 1999, ApJ, to be published

\reference{}
Bulik, T., \& Pavlov, G.~G. 1996, ApJ, 469, 373

\reference{} 
Elsner, R.~F., et al. 1997, AAS Meeting 190, \#09.11

\reference{}
Ginzburg, V.~L., \& Ozernoy, L.~M. 1965, Sov.~Phys. JETP, 20, 689

\reference{}
Gnedin, Yu.~N., \& Pavlov, G.~G. 1974, Sov.~Phys. JETP, 38, 903

\reference{}
Kaaret, P., et al.~1990, Opt. Eng., 29(7), 773

\reference{}
Kaminker, A.~D., Pavlov, G.~G., \& Shibanov, Yu.~A. 1982, 
Ap\&SS, 86, 249 

\reference{}
Lyne, A.~G., \& Manchester, R.~N. 1988, MNRAS, 234, 477 

\reference{}
Manchester, R.~N., \& Taylor, J.~H. 1977, Pulsars (San Francisco: Freeman)

\reference{}
Marshall, H.~L., Murray, S.~S., Silver, E., Schnopper, H., \& Weisskopf, M.~C.
1998, AAS Meeting 192, \#35.04

\reference{}
M\'esz\'aros, P., Novick,~R., Chanan, G.~A., Weisskopf, M.~C.,
\& Szentgyorgyi, A. 1988, ApJ, 324, 1056

\reference{}
\"Ogelman, H. 1995, in The Lives of the Neutron Stars,
ed. M.~A.~Alpar, \"U.~Kizilo\u{g}lu,
\& J.~van Paradijs (Dordrecht: Kluwer), 101

\reference{}
Pavlov, G.~G., \& Shibanov, Yu.~A. 1978, Sov.~Astron., 22, 214

\reference{}
Pavlov, G.~G., \& Shibanov, Yu.~A. 1979, Sov.~Phys.~JETP, 49, 741

\reference{}
Pavlov, G.~G., \& Zavlin, V.~E. 1997, ApJ, 490, L91

\reference{}
Pavlov, G.~G., \& Shibanov, Yu.~A., \& Yakovlev, D.~G. 1980,
Ap\&SS, 73, 33

\reference{}
Pavlov, G.~G., Shibanov, Yu.~A., Ventura, J., \& Zavlin, V.~E. 1994,
A\&A, 289, 837

\reference{}
Pavlov, G.~G., Shibanov, Yu.~A., Zavlin, V.~E., \& Meyer, R.~D. 1995,
in The Lives of the Neutron Stars, ed. M.~A.~Alpar, \"U.~Kizilo\u{g}lu,
\& J.~van Paradijs (Dordrecht: Kluwer), 71

\reference{}
Pavlov, G.~G., Zavlin, V.~E., Tr\"umper, J., \& Neuh\"auser, R. 1996,
ApJ, 472, L33

\reference{}
Pavlov, G.~G., Welty, A.~D., \& C\'ordova, F.~A. 1997, ApJ, 489, L75

\reference{}
Pavlov, G.~G., Zavlin, V.~E., \& Tr\"umper, J. 1999, ApJ, 511, L45

\reference{}
Pineault, S. 1977, MNRAS, 179, 691

\reference{}
Shapiro, S., \& Teukolsky, S. 1983, Black Holes, White Dwarfs
and Neutron Stars (New York: Wiley)

\reference{}
Shibanov, Yu.~A., \& Zavlin, V.~E. 1995, Astron.~Lett., 21, 3

\reference{}
Shibanov, Yu.~A., Zavlin, V.~E., Pavlov, G.~G., Ventura, J.,
\& Potekhin, A.~Y. 1993, in Isolated Pulsars, Eds. K.~A.~Van Riper,
R.~I.~Epstein, \& C.~Ho (Cambridge Univ.~Press: Cambridge)

\reference{}
Tomsick, J.~A., et al. 1997, SPIE, 3114, 373

\reference{}
Walter, F.~M., Wolk, S.~J., \& Neuh\"auser, R. 1996,
Nature, 379, 233

\reference{}
Weisskopf, M.~C., Silver, E.~H., Kestenbaum, K.~S., Long, K.~S.,
Novick, R., \& Wolff, R.~S. 1978, ApJ, 220, L117

\reference{}
Weisskopf, M.~C., Elsner, R.~F., Joy, M.~K., O'Dell, S.~L., Ramsey, B.~D.,
Garmire, G.~P., M\'esz\'aros, P., \& Sunyaev, R.~A. 1994, SPIE Proc,, 2283, 70

\reference{}
Zavlin, V.~E., Shibanov, Yu.~A., \& Pavlov, G.~G. 1995, Astron.~Lett.,
21, 168

\reference{}
Zavlin, V.~E., Tr\"umper, J., \& Pavlov, G.~G. 1999, ApJ, in press

\end{references}
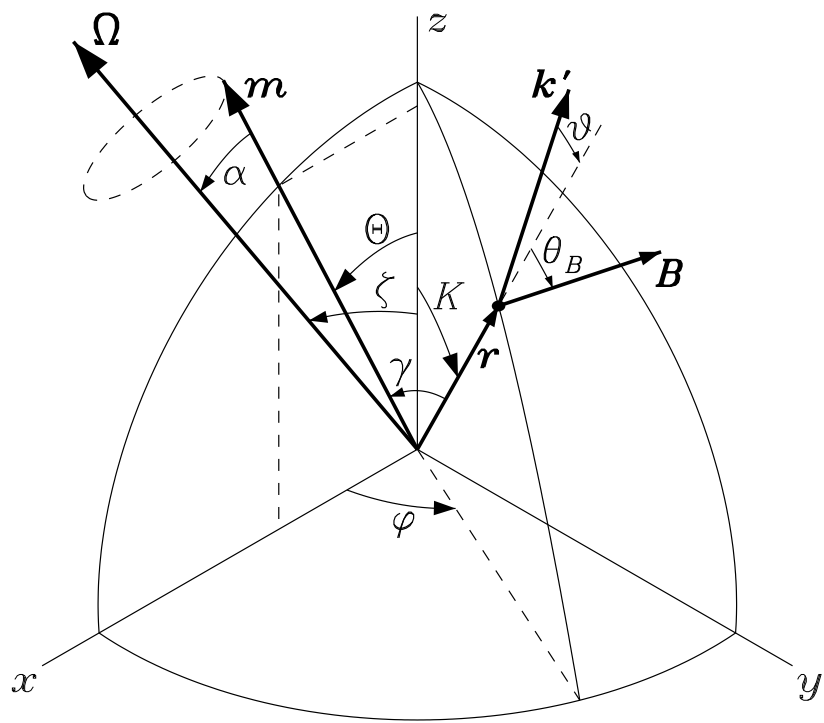
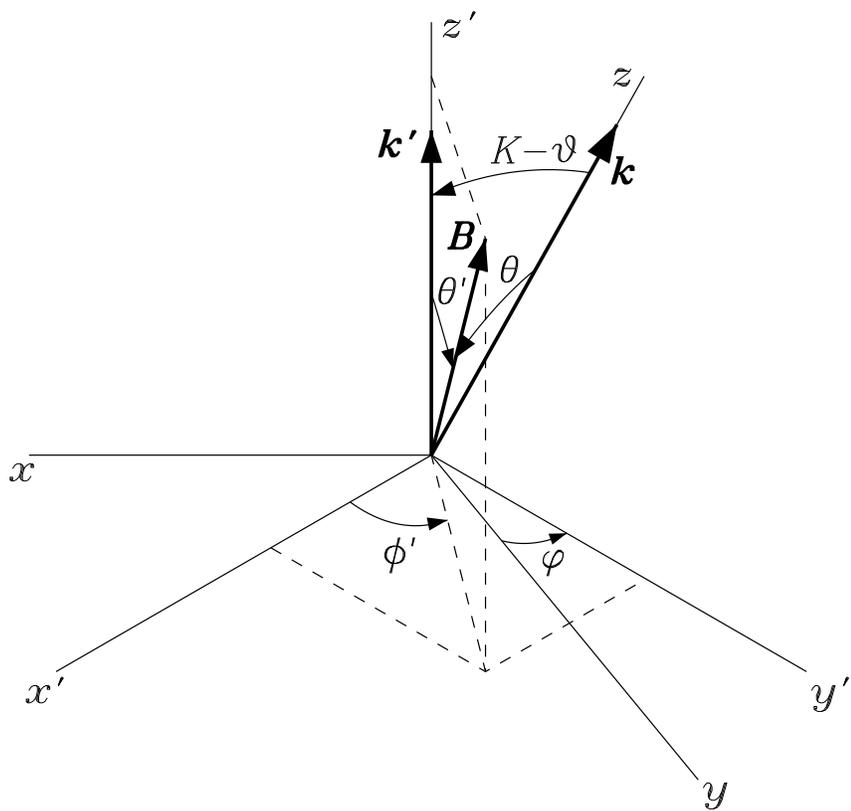
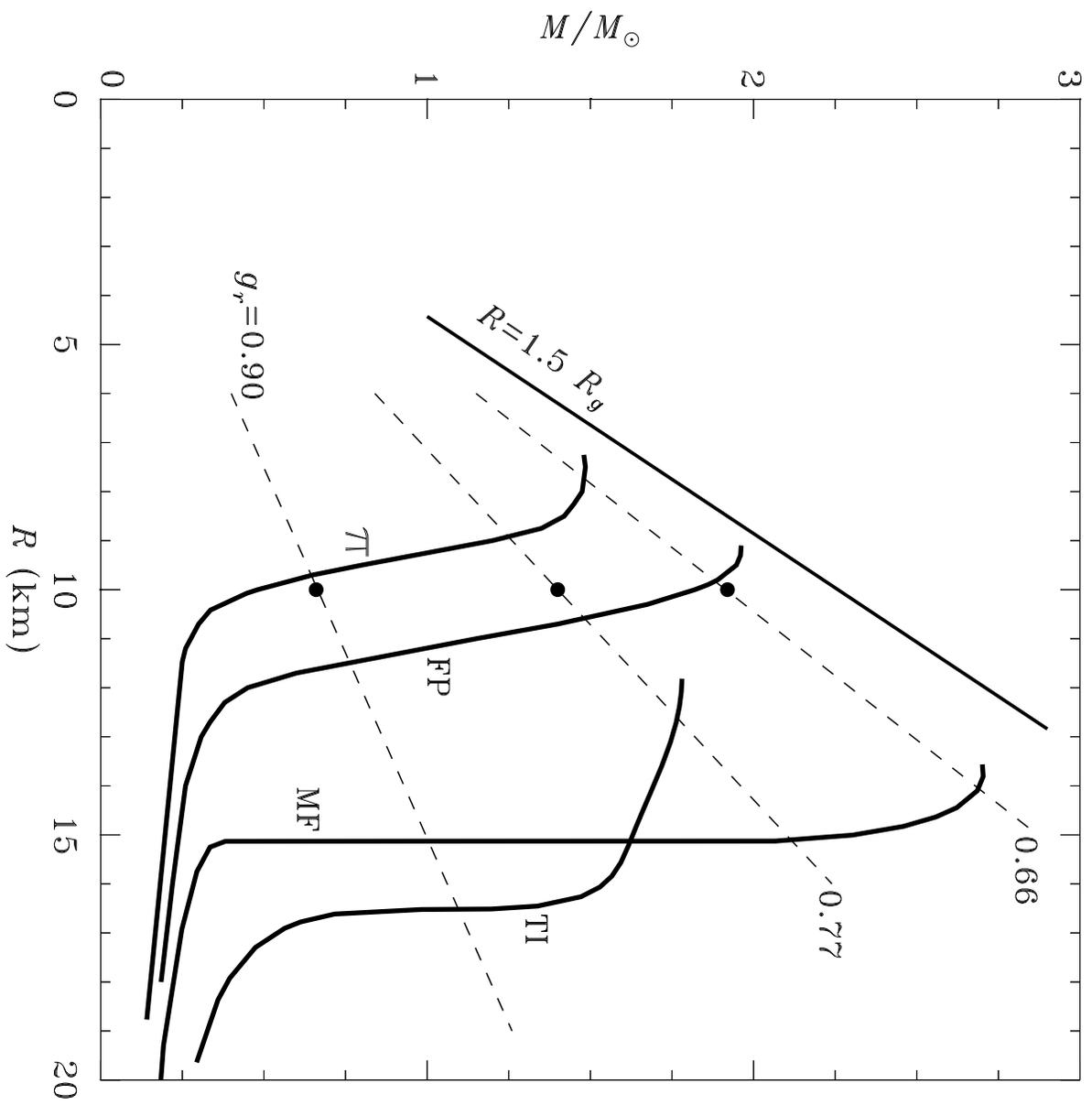
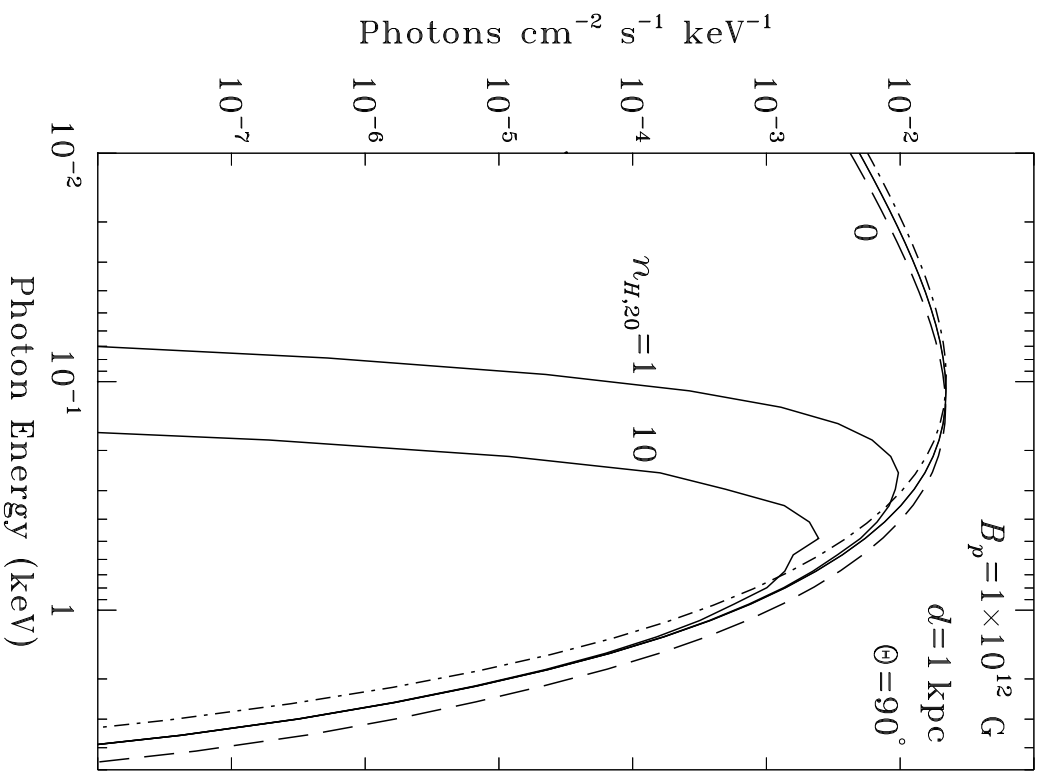
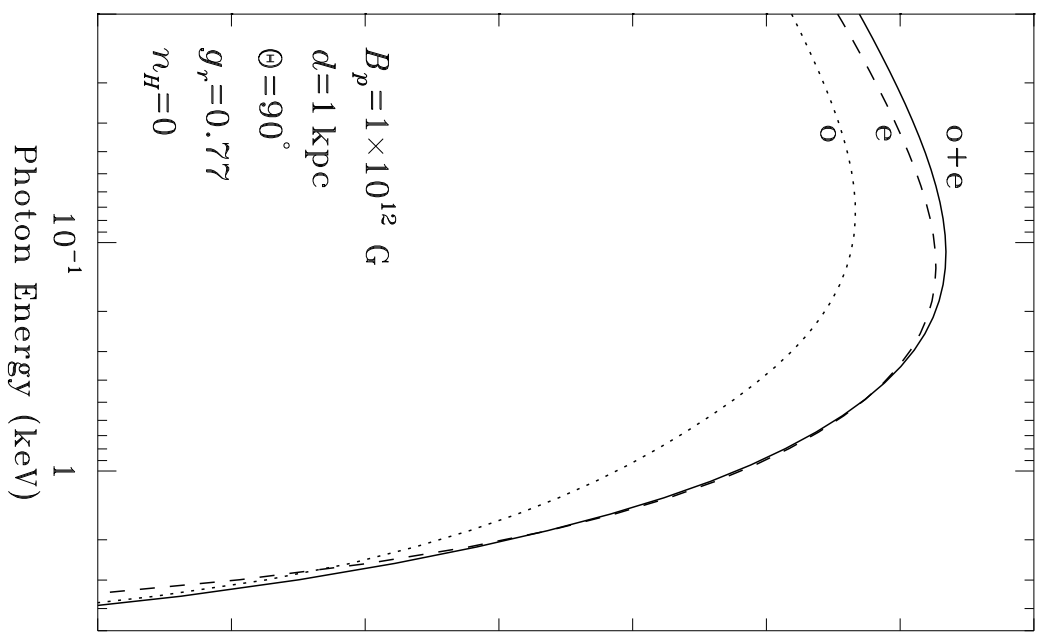
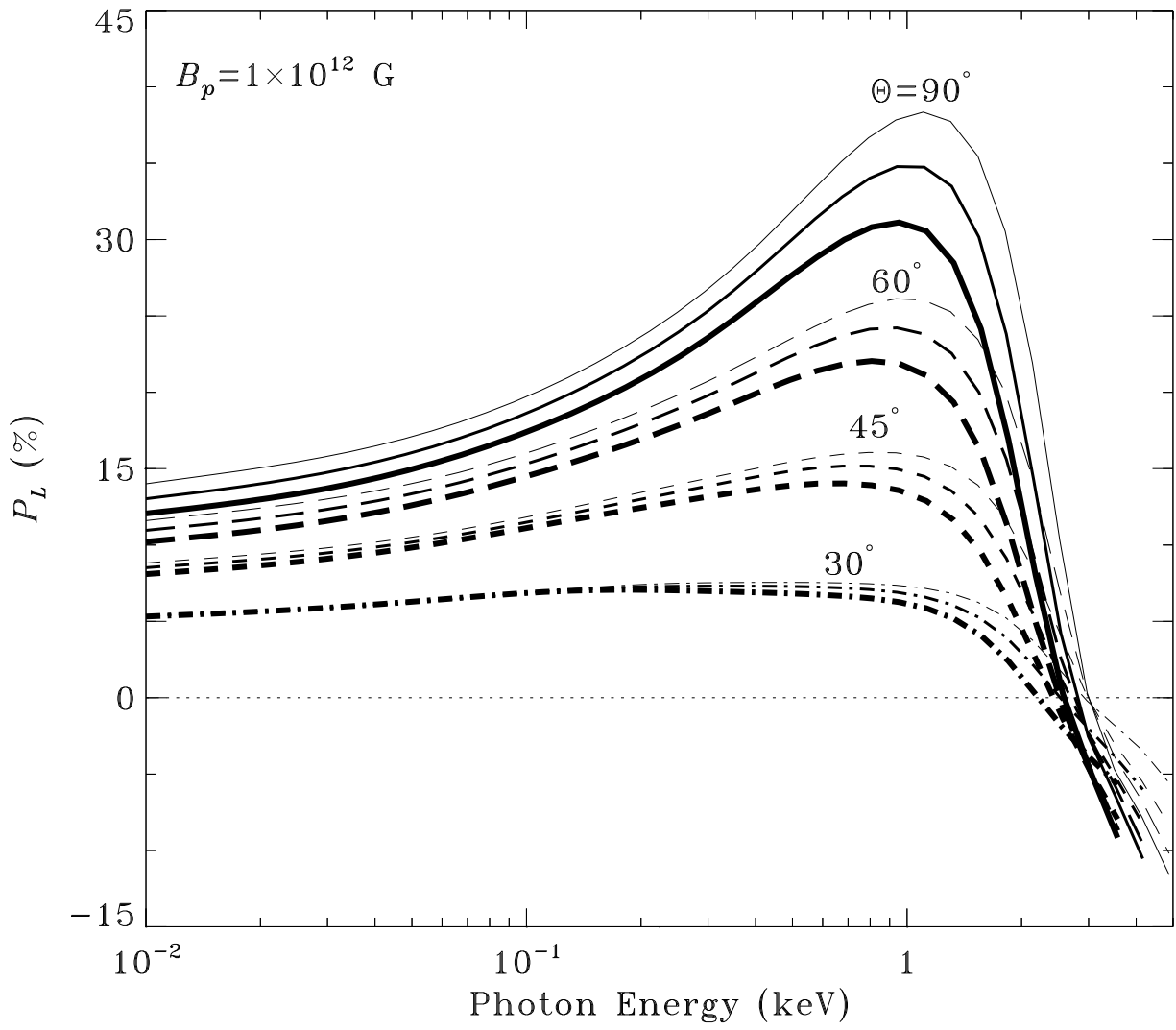
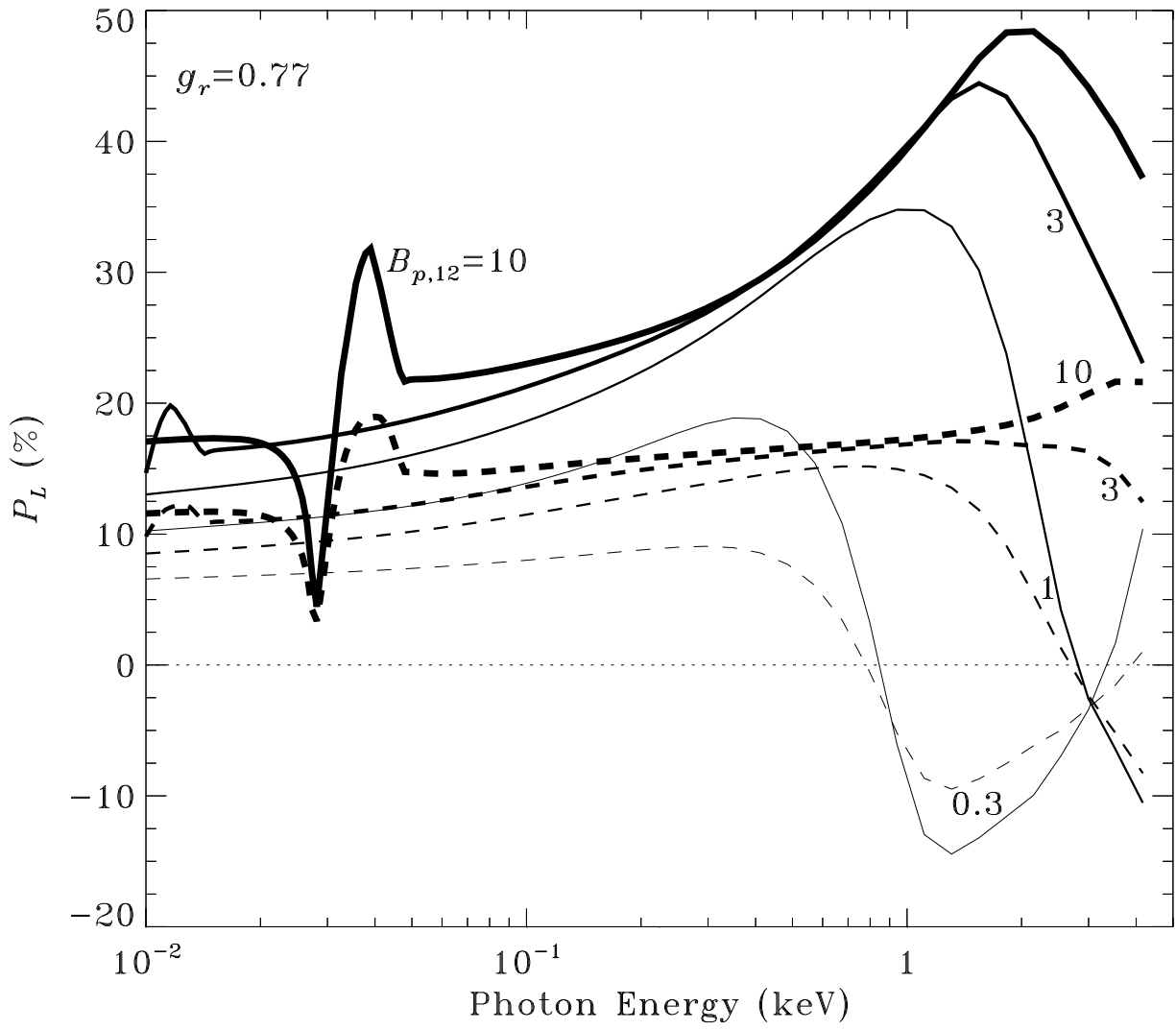
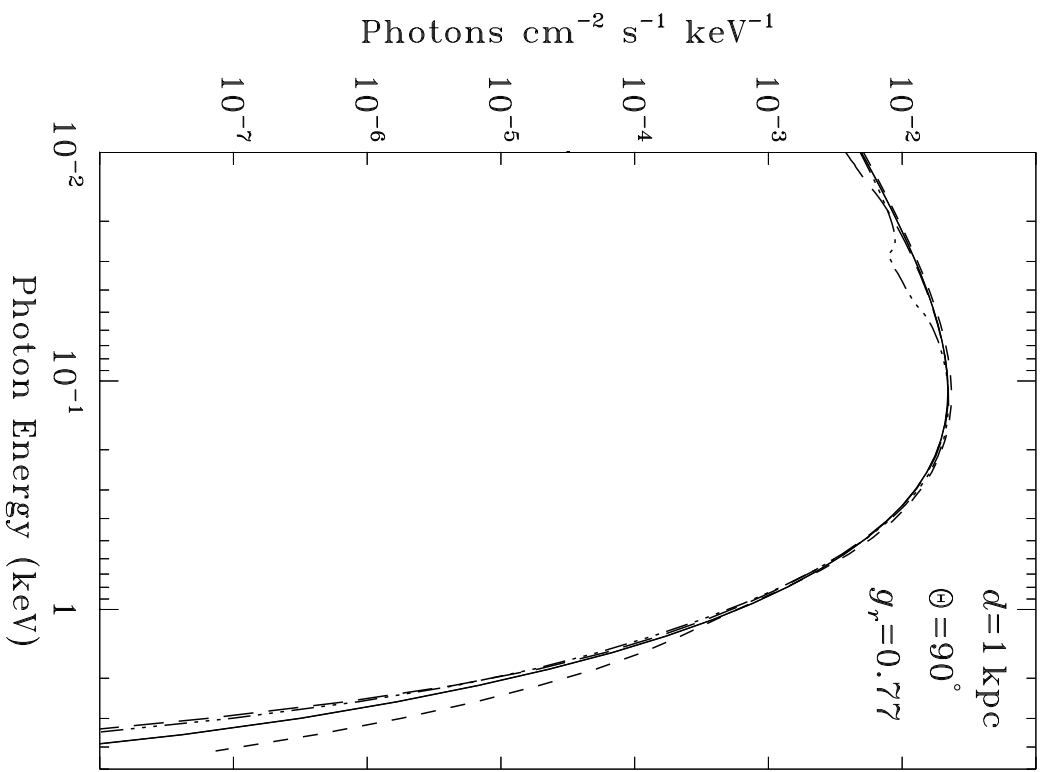
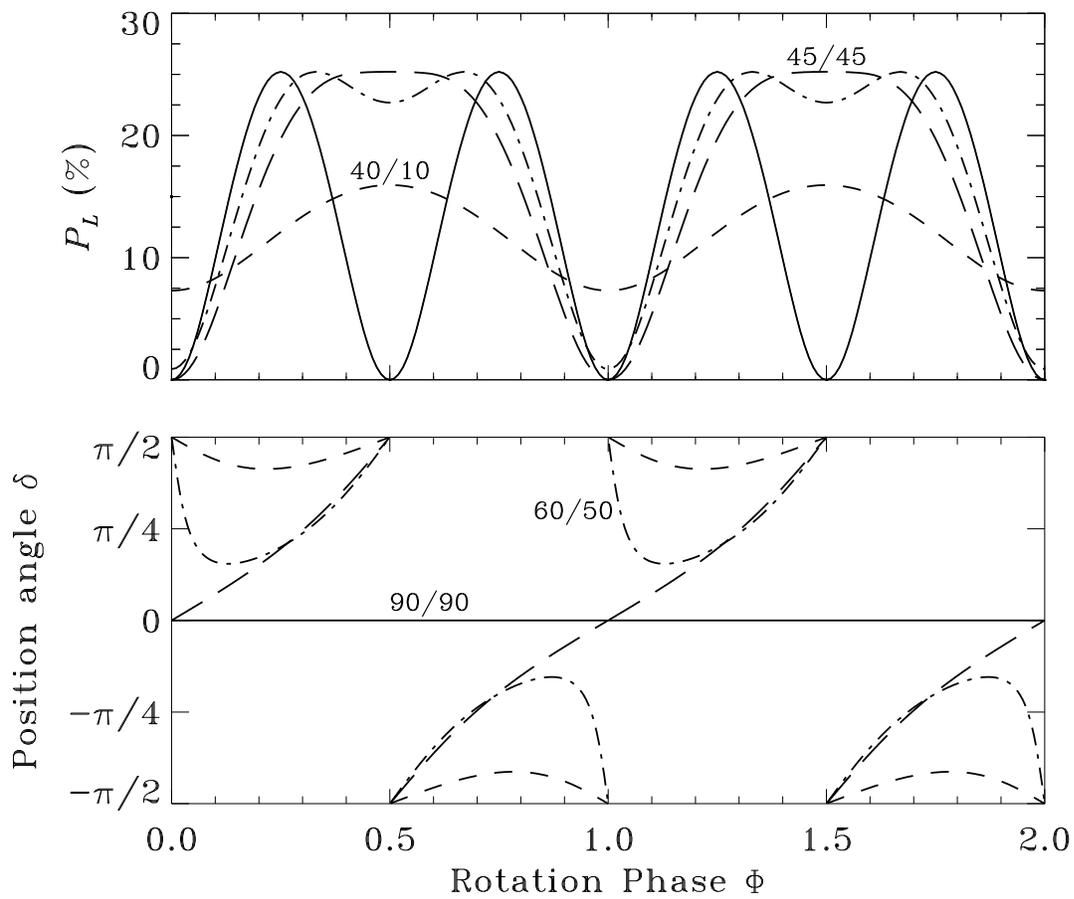
\end{document}